\documentclass[aps,prd,preprint,nofootinbib,11pt]{revtex4}
\usepackage{amsfonts}
\usepackage{mathrsfs}
\usepackage{graphicx}
\usepackage{amsmath}
\usepackage{amssymb}
\usepackage{subfigure}
\usepackage{epsfig}
\usepackage{graphicx}
\usepackage{color}
\newcommand{\bqa}{\begin{eqnarray}}
\newcommand{\eqa}{\end{eqnarray}}
\newcommand{\beq}{\begin{equation}}
\newcommand{\eeq}{\end{equation}}
\allowdisplaybreaks[1]
\graphicspath{{fig/}{dia/}} \DeclareGraphicsExtensions{.eps}
\begin{document}

\title{Scalar Fully-heavy Tetraquark States $QQ^\prime \bar{Q} \bar{Q^\prime}$ in QCD Sum Rules\\[0.7cm]}

\author{Bo-Cheng Yang$^{1}$, Liang Tang$^{1, 2}$\footnote{tangl@hebtu.edu.cn}, and Cong-Feng Qiao$^{3,4}\footnote{qiaocf@ucas.ac.cn}$}

\affiliation{$^1$ College of Physics, Hebei Normal University, Shijiazhuang 050024, China\\
$^2$ Hebei Key Laboratory of Photophysics Research and Application, Hebei Normal University, Shijiazhuang 050024, China\\
$^3$ School of Physics, University of CAS, Beijing 100049, China\\
$^4$ CAS Center for Excellence in Particle Physics, Beijing 100049, China
}



\begin{abstract}
\vspace{0.3cm}
Very recently, the LHCb Collaboration observed distinct structures with the $cc\bar{c}\bar{c}$ in the $J/\Psi$-pair mass spectrum. In this work, we construct four scalar ($J^{PC} = 0^{++}$) $[8_c]_{Q\bar{Q^\prime}}\otimes [8_c]_{Q^\prime \bar{Q}}$ type currents to investigate the fully-heavy tetraquark state $Q Q^\prime \bar{Q} \bar{Q^\prime}$ in the framework of QCD sum rules, where $Q=c, b$ and $Q^\prime = c, b$. Our results suggest that the broad structure around 6.2-6.8 GeV can be interpreted as the $0^{++}$ octet-octet tetraquark states with masses $6.44\pm 0.11$ GeV and $6.52\pm 0.10$ GeV, and the narrow structure around $6.9$ GeV can be interpreted as the $0^{++}$ octet-octet tetraquark states with masses $6.87\pm 0.11$ GeV and $6.96\pm 0.11$ GeV, respectivley. Extending to the b-quark sector,the masses of their fully-bottom partners are found to be around 18.38-18.59 GeV. Additionally, we also analyze the spectra of the $[8_c]_{c\bar{c}}\otimes [8_c]_{b \bar{b}}$ and $[8_c]_{c\bar{b}}\otimes [8_c]_{b \bar{c}}$ tetraquark states, which lie in the range of 12.51-12.74 GeV and 12.49-12.81 GeV, respectively.

\end{abstract}
 \maketitle
\newpage

\section{Introduction}

Hadrons with more than the minimal quark content ($q\bar{q}$ or $qqq$) was proposed by Gell-Mann \cite{GellMann:1964nj} and Zweig \cite{Zweig} in 1964, which were named as multiquark exotic states and were also allowed by Quantum Chromodynamics (QCD). Since the discovery of X(3872) in 2003~\cite{Choi:2003ue}, benefitted from the accumulation of more and more experimental data, many multiquark states have been discovered at B factories, LHC and BESIII experiments. Needless to say, understanding the nature of the multiquark exotic states is one of the most intriguing research topics of hadronic physics, which attracts more and more interests from theorists and experimentalists.

Very recently, the LHCb Collaboration reported resonant structures in the double-$J/\psi$ invariant mass distribution using data for $pp$ collisions at centre-of-mass energies of 7, 8 and 13 TeV collected by the LHCb experiment at the Large Hadron Collider, corresponding to an integrated luminosity of $9 \text{fb}^{-1}$~\cite{Aaij:2020fnh}. They observed a broad structure above the threshold ranging from 6.2 to 6.8 GeV and a narrow structure at around 6.9 GeV, referred to as X(6900). Such structures are naturally assigned to have the constituent quarks $c c \bar{c} \bar{c}$, making them the first fully-heavy multi-quark exotic candidates claimed to date in the experimental literature. In the fully-bottom tetraquark sector, the CMS Collaboration observed the $\Upsilon(1S)$ pair production and indicated a $bb\bar{b} \bar{b}$ signal around 18.4 GeV with a global significance of $3.6 \, \sigma$~\cite{Khachatryan:2016ydm} in 2017. However, the LHCb  did not obeserved this tetraquark state by searching the invariant mass distribution of $\Upsilon(1S) \mu^+ \mu^-$~\cite{Aaij:2018zrb, Sirunyan:2020txn}. The contradictory information from CMS and LHCb Collaborations on fully-bottom tetraquark state urge more researches on its correspording fully-charm tetraquark partner for both theorists and experimentalists.

Theoretical studies of $c c \bar c\bar{c}$ states have a much longer history~\cite{Iwasaki:1976cn}. As early as in 1981, the exotic hadrons composed of $c c \bar{c} \bar{c}$ have been systematically studied in quark-gluon model by Chao for the first time\cite{Chao:1980dv}, where their masses were predicted to lie in the range 6.4-6.8 GeV and were all above the threshold of two charmonia. Later, the fully-charm tetraquark states were investigated in Refs.~\cite{Ader:1981db, Badalian:1985es, Heller:1985cb}, which have predicted that fully-charm tetraquark states mainly decay into two charmonia. In recent years, the fully heavy tetraquark states attracted much attention and have been studied extensively in various model schemes~\cite{Anwar:2017toa, Bedolla:2019zwg, Dong:2020nwy, Lloyd:2003yc, Barnea:2006sd, Debastiani:2017msn, Wu:2016vtq, Wang:2019rdo, Liu:2019zuc, Faustov:2020qfm, Lu:2020cns, Heupel:2012ua, Weng:2020jao, Berezhnoy:2011xn, Karliner:2016zzc, Chao:2020dml, Maiani:2020pur, Richard:2020hdw, Berezhnoy:2011xy, Feng:2020riv, Ma:2020kwb, Karliner:2020dta, Wang:2020wrp, Giron:2020wpx, Maciula:2020wri, Zhu:2020xni, Guo:2020pvt, Zhu:2020snb, Eichmann:2020oqt, Gong:2020bmg, Becchi:2020uvq, Chen:2016jxd, Wang:2017jtz, Chen:2018cqz, Wang:2018poa, Zhang:2020xtb, Wang:2020dlo, Albuquerque:2020hio, Chen:2020xwe, Wan:2020fsk}. Among these methods, the QCD sum rules technique has some peculiar advantages in exploring hadron properties involving nonpertubative
QCD, and has been used to analyze X(6900) in recent works~\cite{Chen:2016jxd, Wang:2017jtz, Chen:2018cqz, Wang:2018poa, Zhang:2020xtb, Wang:2020dlo, Albuquerque:2020hio, Chen:2020xwe, Wan:2020fsk}.

Since the broad structure and X(6900) respectivley lie well above the $\eta_c \eta_c$ threshold and $J/\psi J/\psi$ threshold and they don't contain light flavor quark, the four-charm structures are unlikely to be the hadronic molecules, which are usually formed by light meson exchanges with small binding energies. Then the color binding diquark-antidiquark structure is widely used to interpret the newly reported X(6900). It should be noted that according to Quantum Chromodynamics (QCD), there exists another possible tetraquark configuration $[8_c]_{Q\bar{Q}}\otimes [8_c]_{Q \bar{Q}}$, which is composed of two color-octet parts~\cite{Latorre:1985uy, Narison:1986vw, Wang:2006ri, Wang:2015nwa, Tang:2016pcf, Tang:2019nwv}. Since there exists a QCD interaction, it is different from molecular state with two color-singlet mesons. That is to say, it could decay to two charmonia via exchanging one or more gluons. Therefore, the study of the color-octet tetraquark state is very important for possible new exotic hadrons. In this paper, we endeavor to analyze in the framework of QCD sum rules whether there exists stable $J^{PC} = 0^{++}$ tetraquark states with color octet-octet configuration or not, and compare the results with the above-mentioned new structures reported by the LHCb Collaboratoin.

The rest of the paper is arranged as follows. After the introduction, some primary formulas of the QCD sum rules in our calculation are presented in Sec. \ref{Formalism}. The numerical analysis and results are given in Sec. \ref{Numerical}. The last part is left for conclusions and discussion of the results.

\section{Formalism}\label{Formalism}

The starting point of the QCD sum rules~\cite{Shifman, Reinders:1984sr, Narison:1989aq, P.Col} is the two-point correlation function constructed from two hadronic currents. For a scalar state considered in this work, the two-point correlation function is expressed as the following form:
\begin{eqnarray}
\Pi(q) &=& i \int d^4 x e^{i q \cdot x} \langle 0 | T \{ j(x), j^\dagger(0) \} |0 \rangle \; ,
\end{eqnarray}
where $j(x)$ and $j(0)$ are the interpolating currents with $J^{PC} = 0^{++}$.

The interpolating currents of fully-charm or fully-bottom tetraquark states with $J^{PC} = 0^{++}$ are, respectively, constructed as
\begin{eqnarray}
j_{A}^{0^{++}}(x) &=& [i\overline{Q}^j(x)\gamma^5(t^a)_{jk}Q^k(x)][i\overline{Q}^m(x)\gamma^5(t^a)_{mn}Q^n(x)]  \; \label{current-1} , \\
j_{B}^{0^{++}}(x) &=& [\overline{Q}^{j}(x) \gamma_{\mu} (t^{a})_{jk} Q^{k}(x)][\overline{Q}^{m}(x) \gamma^{\mu} (t^{a})_{mn} Q^{n}(x)] \; \label{current-2}  , \\
j_{C}^{0^{++}}(x) &=& [i\overline{Q}^j(x)(t^a)_{jk}Q^k(x)][i\overline{Q}^m(x)(t^a)_{mn}Q^n(x)] \; \label{current-3} , \\
j_{D}^{0^{++}}(x) &=& [\overline{Q}^{j}(x) \gamma_{\mu} \gamma_{5} (t^{a})_{jk} Q^{k}(x)][\overline{Q}^{m}(x) \gamma^{\mu} \gamma_{5} (t^{a})_{mn} Q^{n}(x)]\; \label{current-4}\  ,
\end{eqnarray}
where $j, k, m$, and $n$ are the color indices, the $t^a$ is the Gell-Mann matrix, and $Q$ represents the heavy-quark $c$ or $b$. Here, the subscripts $A$ to $D$ of $J^{PC} = 0^{++}$ indicate the currents composed of two $0^-$ color-octet $Q\bar{Q}$ system, two $1^-$ color-octet $Q \bar{Q}$ system, two $0^+$ color-octet $Q \bar{Q}$ system,  and two $1^+$ color-octet $Q \bar{Q}$ system, respectively.

At the quark-gluon level, the correlation function can be calculated with the operator product expansion (OPE). In our evaluation, the heavy-quark ($Q=c$ or $b$) propagator $S^Q_{ij}(p)$ is considered in momentum space, which can be expanded as
\begin{eqnarray}
S^Q_{j k}(p) \! \! & = & \! \! \frac{i \delta_{j k}(p\!\!\!\slash + m_Q)}{p^2 - m_Q^2} - \frac{i}{4} \frac{t^a_{j k} G^a_{\alpha\beta} }{(p^2 - m_Q^2)^2} [\sigma^{\alpha \beta}
(p\!\!\!\slash + m_Q)
+ (p\!\!\!\slash + m_Q) \sigma^{\alpha \beta}] \nonumber \\ &+& \frac{i\delta_{jk}m_Q  \langle g_s^2 G^2\rangle}{12(p^2 - m_Q^2)^3}\bigg[ 1 + \frac{m_Q (p\!\!\!\slash + m_Q)}{p^2 - m_Q^2} \bigg] \nonumber \\ &+& \frac{i \delta_{j k}}{48} \bigg\{ \frac{(p\!\!\!\slash +
m_Q) [p\!\!\!\slash (p^2 - 3 m_Q^2) + 2 m_Q (2 p^2 - m_Q^2)] }{(p^2 - m_Q^2)^6}
\times (p\!\!\!\slash + m_Q)\bigg\} \langle g_s^3 G^3 \rangle \; .
\end{eqnarray}
Here, the vacuum condensates are clearly displayed. For more interpretation on above propagators, the reader is referred to Refs.~\cite{Wang:2013vex, Albuquerque:2013ija}.

Based on the dispersion relation, the correlation function $\Pi(q^2)$ of the quark -gluon side can be expressed as
\begin{eqnarray}
  \Pi^{OPE}(q^2) = \int_{(4 m_Q)^2}^\infty ds \frac{\rho^{OPE}(s)}{s - q^2} + \Pi^{\langle GG \rangle}(q^2), \label{Pi-OPE}
\end{eqnarray}
where $\rho^{OPE}(s) = \text{Im} [\Pi^{OPE}(s)]/\pi$ and
\begin{eqnarray}
  \rho^{OPE}(s) &=& \rho^{\text{pert}}(s) + \rho^{\langle GG \rangle}(s).
\end{eqnarray}
The second term $\Pi^{\langle GG \rangle}(q^2)$ in Eq.(\ref{Pi-OPE}) represents the contribution in the correlation function that have no imaginary part but have nontrivial magnitudes after the Borel transformation. After making the Borel transformation to Eq.(\ref{Pi-OPE}), we can obtain
\begin{eqnarray}
  \Pi^{\text{OPE}}(M_B^2) = \int_{(4 m_Q)^2}^\infty ds \rho^{\text{OPE}}(s) e^{-s/M_B^2} + \Pi^{\langle GG \rangle}(M_B^2). \label{Pi-MB}
\end{eqnarray}

For all the tetraquark states considered in this work, we put the lengthy expressions of spectral densities $\rho^{OPE}(s)$ and $\Pi^{\langle GG \rangle} (M_B^2)$ in Eq.(\ref{Pi-MB}) into the Appendix.

On the phenomenological side, after separating the ground state contribution from the pole term, the correlation function $\Pi(q^2)$ can be expressed as the dispersion integral over the physical region, \textit{i.e.},
\begin{eqnarray}
  \Pi(q^2) = \frac{(\lambda_X)^2}{(M_X)^2 - q^2} + \frac{1}{\pi} \int_{s_0}^\infty ds \frac{\rho(s)}{s - q^2}, \label{Pi-hadron}
\end{eqnarray}
where the subscript $X$ means the lowest lying tetraquark state, $M_X$ denotes its mass, and $\rho(s)$ is the spectral density that contains the contribution from higher excited states and the continuum states above the threshold $s_0$. As in Refs.\cite{Reinders:1984sr,P.Col}, in order to work out the phenomenological side of the QCD sum rules, a complete set of intermediate states should be inserted into the two color-octet-octet tetraquark interpolating currents, where the summation goes over all possible hadronic states created by the color-octet-octet tetraquark current. The coupling constant $\lambda_X$ is defined through
\begin{eqnarray}
\langle 0|j^{0^{++}}(0)|X\rangle &=& \lambda_X,
\end{eqnarray}
where $X$ stands for the tetraquark state.

By performing the Borel transform on the phenomenological side, Eq.(\ref{Pi-hadron}), and matching it to Eq.(\ref{Pi-MB}), we can then obtain the mass of the tetraquark state,
\begin{eqnarray}
  M_X^{i}(s_0, M_B^2) &=& \sqrt{-\frac{L_1(s_0, M_B^2)}{L_0(s_0, M_B^2)}}, \label{main-function}
\end{eqnarray}
where the superscript $i$ runs from $A$ to $D$, respectively. The moments $L_1$ and $L_0$ are, respectively, defined as
\begin{eqnarray}
  L_0(s_0, M_B^2) &=& \int_{(4 m_Q)^2}^\infty ds \rho^{OPE}(s) e^{-s/M_B^2} + \Pi^{\langle GG \rangle}(M_B^2), \label{L0} \\
  L_1(s_0, M_B^2) &=& \frac{\partial}{\partial (M_B^2)^{-1}} L_0(s_0, M_B^2).
\end{eqnarray}

\section{Numerical Evaluation}\label{Numerical}

In order to yield meaningful physical results in QCD sum rules, as in any practical theory, one needs to give certain inputs . To perform numerical analyses, we use the following values for various condensate and heavy quark masses~\cite{Shifman, Reinders:1984sr, P.Col, Narison:1989aq}:
$m_c (m_c) = \overline{m}_c= (1.27 \pm 0.03) \; \text{GeV}$,
$m_b (m_b) = \overline{m}_b= (4.18 \pm 0.03) \; \text{GeV}, $
$\langle g_s^2 G^2 \rangle = 0.48 \pm 0.14 \; \text{GeV}^4$,
where, $\overline{m}_c$ and $\overline{m}_b$ represent heavy-quark running masses in $\overline{MS}$ scheme.

Moreover, there exist two additional parameters $M_B^2$ and $s_0$ introduced in establishing the sum rule, which will be constrained by three criteria\cite{Shifman, Reinders:1984sr, P.Col}.
First, to extract the information on ground state tetraquark state, one should guarantee pole contribution (PC) is bigger than the continuum contribution~\cite{P.Col, Matheus:2006xi}, which can be determined by the formula
\begin{eqnarray}
  R_{i}^{PC} = \frac{L_0(s_0, M_B^2)}{L_0(\infty, M_B^2)} \; , \label{RatioPC}
\end{eqnarray}
where the subscript $i$ runs from $A$ to $D$. Under this prerequisite, the most part of the contribution in the mass equation~(\ref{main-function}) comes directly from the ground state, and the critical value of $M_B^2$ is the upper limit $(M_B^2)_{max}$.

The second one asks for the convergence of the OPE, which fixes the lower limit on $M_B^2$, that is, $(M_B^2)_{min}$. In general, one can determine the $(M_B^2)_{min}$ value by the following ratio
\begin{eqnarray}
  R_{i}^{OPE} = \frac{L_0^{dim}(s_0, M_B^2)}{L_0(s_0, M_B^2)}\, ,
\end{eqnarray}
where the contribution of the higher dimension condensate in the OPE side is smaller than 10\% to 25\%of the total contribution~\cite{Albuquerque:2013ija, P.Col}. Here the superscript dim means the dimension of relevant condensate in the OPE of Eq.(\ref{L0}), the subscript $i$ runs from $A$ to $D$. As in Refs.~\cite{Wang:2017jtz, Wang:2020dlo, Chen:2016jxd, Zhang:2020xtb}, for the fully heavy tetraquark systems, the three-gluon condensate is not only lower than the two-gluon contribution but also too tiny in our calculations.  Then we conclude that if $M_B^2$ are larger than $4.0\, \text{GeV}^2$, the operator product expansion is well convergent for the scalar fully-charm tetraquark states. Therefore, for simplicity, we neglect three-gluon condensate in our numerical analyses in the following.

The third criterion is to require the dependence of the mass of the tetraquark state $M_X$ on the parameter $s_0$ to be weak. To find a proper value for continuum threshold $s_0$, we perform a similar analysis as in Refs.~\cite{Finazzo:2011he, Qiao:2013raa, Qiao:2013dda}. Notice that the $s_0$ relates to the mass of the ground state by $\sqrt{s_0} \sim (M_X + \delta) \, \text{GeV}$, in which $\delta$ lies in the scope of $0.4 \sim 0.8$ GeV. Therefore, various $\sqrt{s_0}$ satisfying this constraint should be taken into account in the numerical analysis. Among these values, we need to pick up the one which yields an optimal window for Borel parameter $M_B^2$. That is to say, in the optimal window, the fully heavy tetraquark mass $M_X$ is somehow independent of the Borel parameter $M_B^2$. Eventually, the value of $\sqrt{s_0}$ corresponding to the optimal mass curve will be taken as its central value. In practice, in order to estimate the uncertainties stemming from $s_0$, we may vary $\sqrt{s_0}$ by $0.20$ GeV in numerical calculation \cite{Qiao:2013raa}, which set the upper and lower bounds on $\sqrt{s_0}$.

\begin{figure}[htb]
\begin{center}
\includegraphics[width=7.0cm]{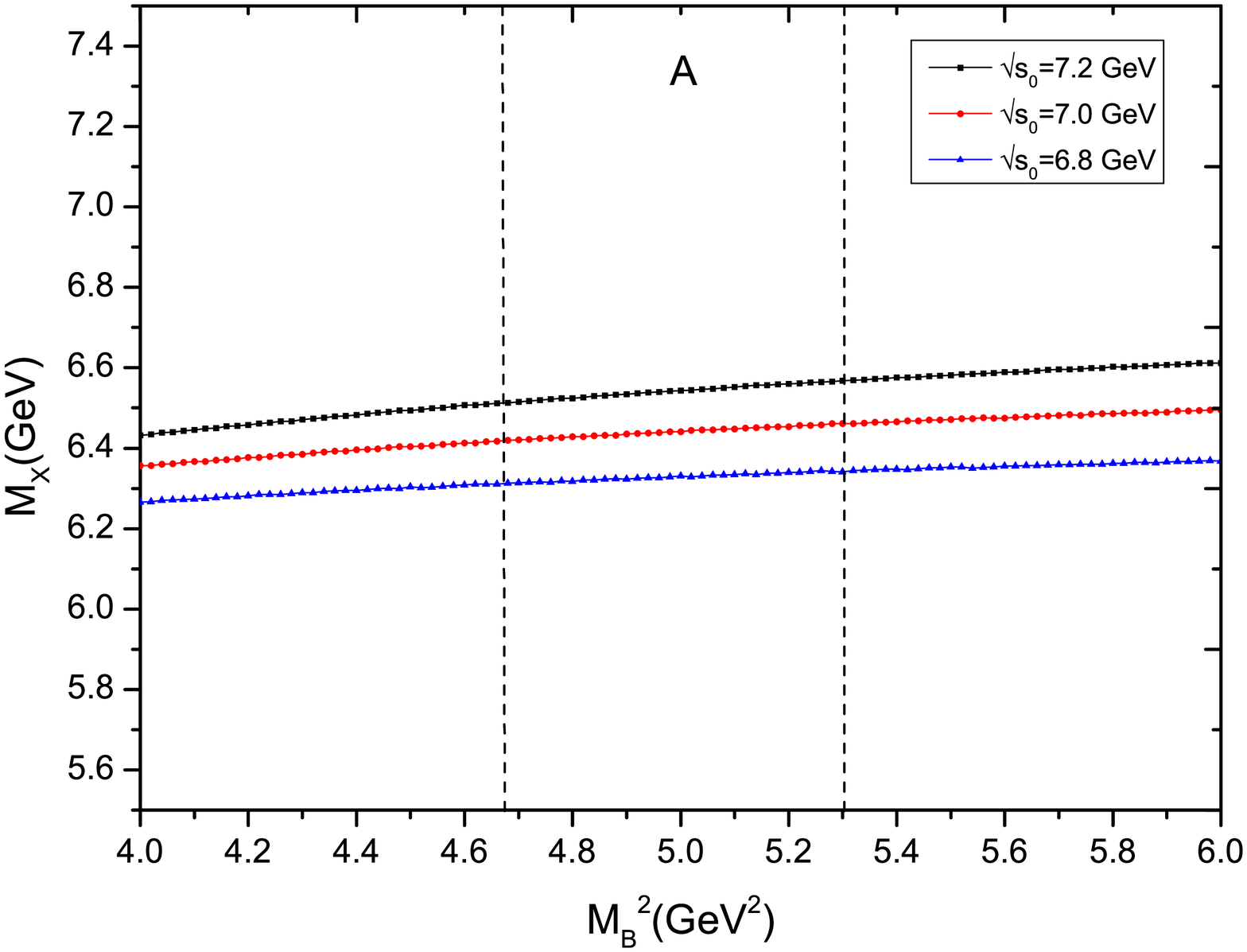}
\includegraphics[width=7.0cm]{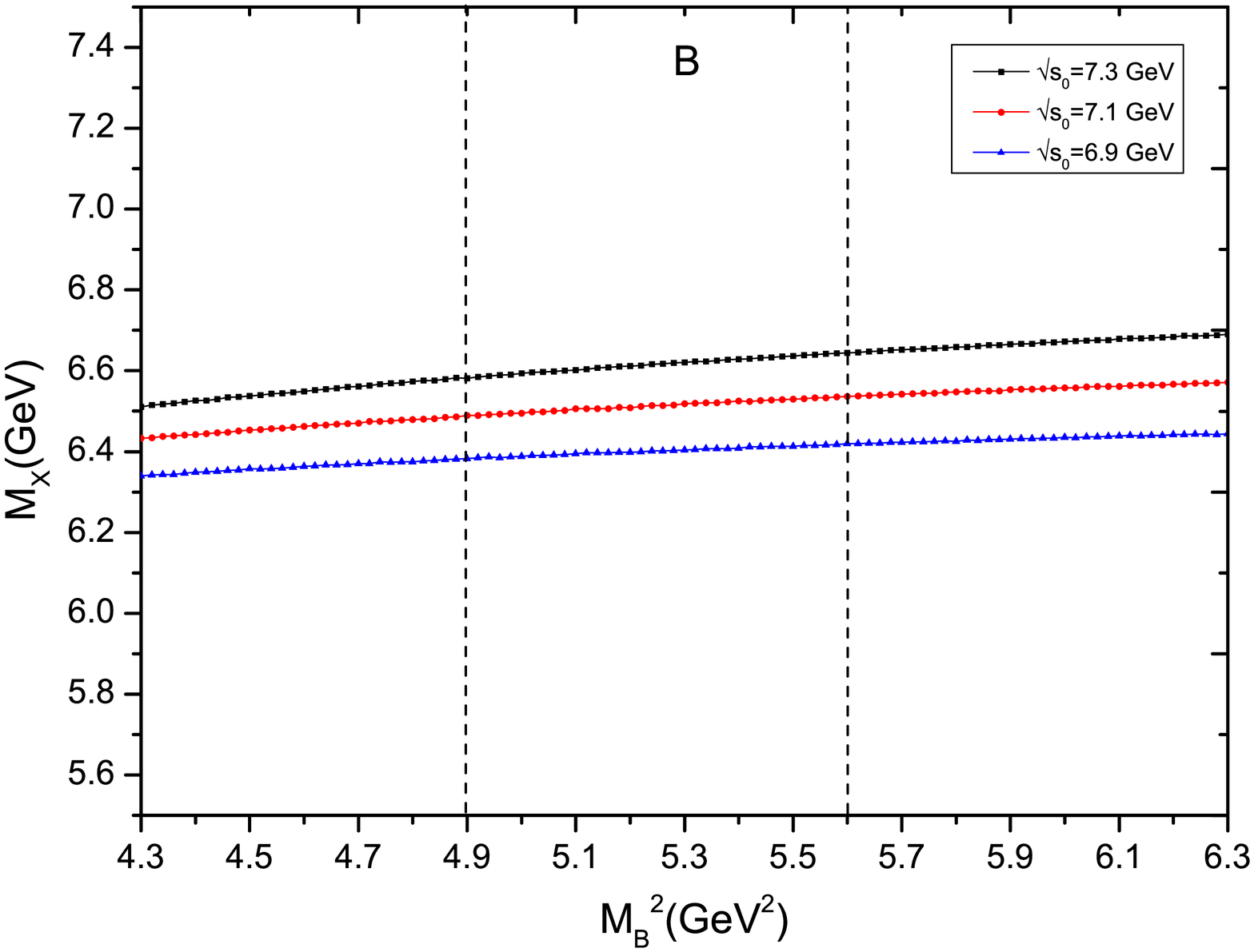}
\includegraphics[width=7.0cm]{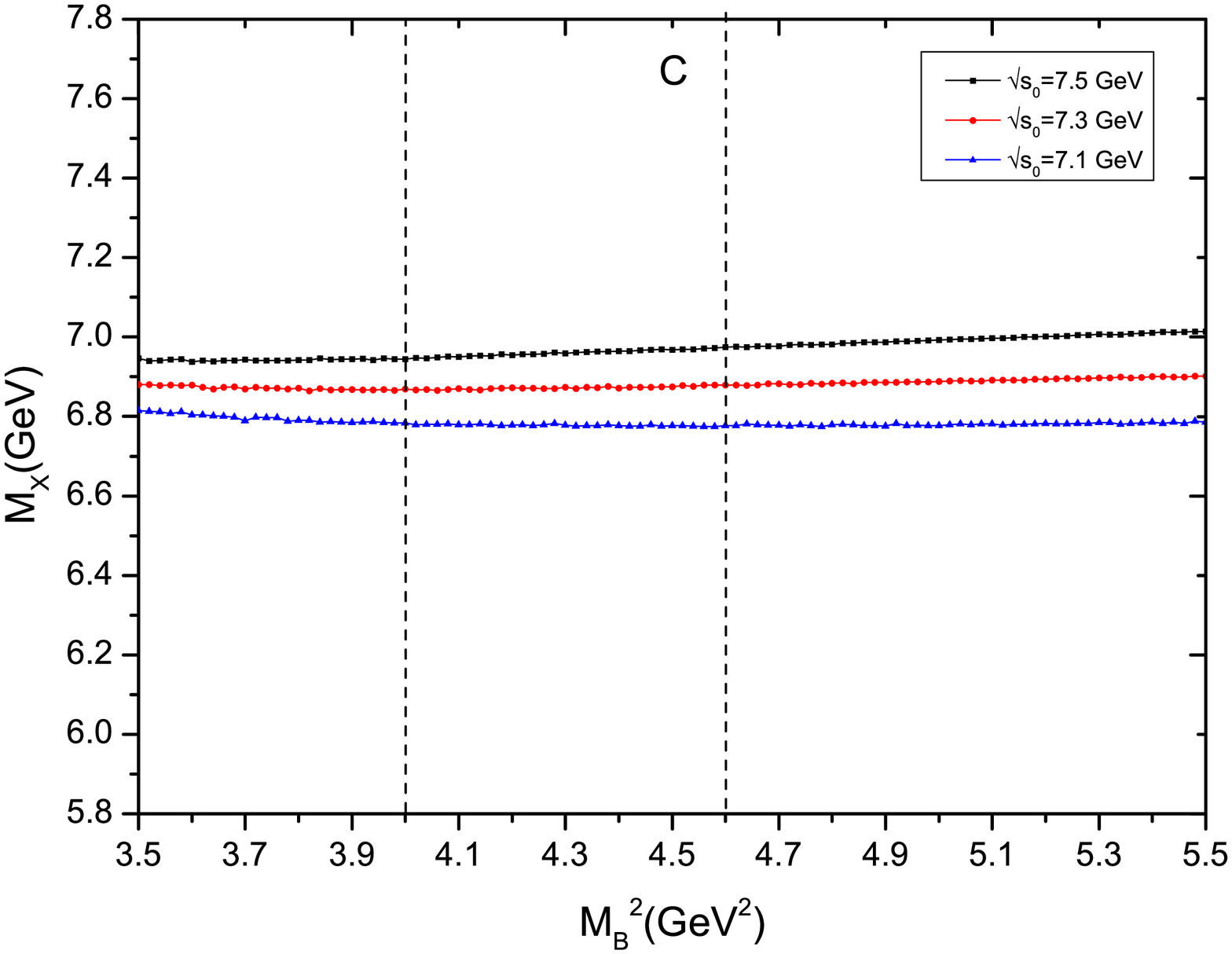}
\includegraphics[width=7.0cm]{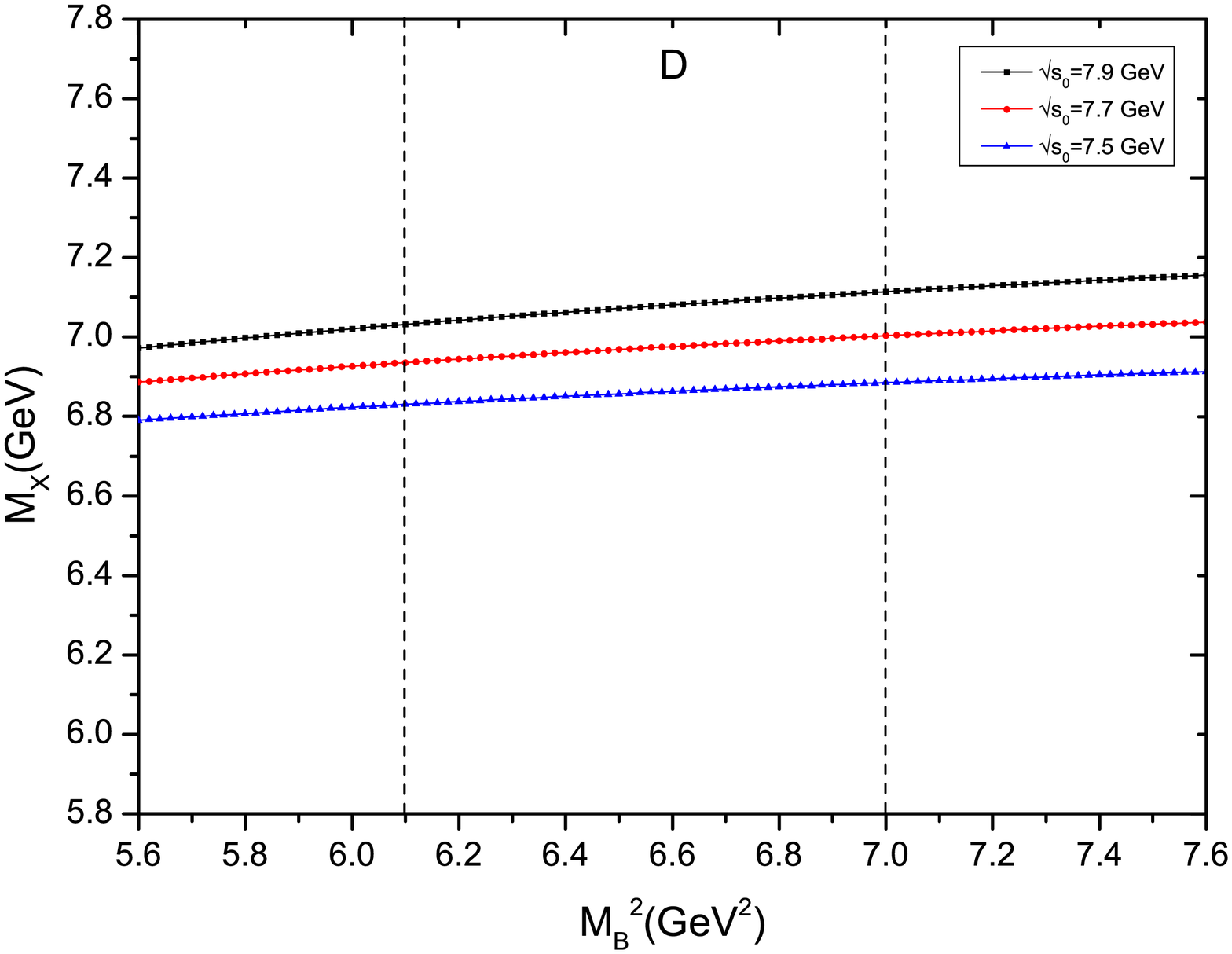}
\caption{The $[8_c]_{c\bar{c}}\otimes [8_c]_{c \bar{c}}$ tetraquark masses $M_X$ as a function of the Borel parameter $M_B^2$ with $J^{PC} = 0^{++}$ and different values of $\sqrt{s_0}$, where $A$ to $D$ stand for cases $A$ to $D$, and the two vertical lines in each case indicate the upper and lower bounds of the valid Borel window with the central value of $\sqrt{s_0}$.} \label{fig1}
\end{center}
\end{figure}

\begin{table}[htb]
\begin{center}
\begin{tabular}{|c|c|c|c|c|c|c}\hline\hline
& $M_B^2 (\rm{GeV}^2)$ & $\sqrt{s_0} (\rm{GeV})$ & PC & $R_{i}^{\langle GG \rangle}$ & $M_X$ (\rm{GeV})  \\ \hline
 $0^{++}$ case $A$ & $4.7-5.3$   & $7.0\pm 0.2$  & $(61-50)\%$ & $(0.37-0.24)\%$  & $6.44 \pm 0.11 $  \\ \hline
 $0^{++}$ case $B$ & $4.9-5.6$   & $7.1\pm 0.2$  & $(60-50)\%$ & $(0.05-0.02)\%$  & $6.52 \pm 0.11 $ \\ \hline
 $0^{++}$ case $C$ & $4.0-4.6$   & $7.3\pm 0.2$   & $(62-50)\%$ & $(28.98-23.60)\%$ & $6.87\pm 0.10 $       \\ \hline
 $0^{++}$ case $D$ & $6.1-7.0$   & $7.7\pm 0.2$   & $(62-50)\%$ & $(2.77-2.48)\%$  & $6.96 \pm 0.11 $    \\ \hline
 \hline
\end{tabular}
\end{center}
\caption{The windows of the Borel parameter, continuum thresholds, pole contributions, two-gluon contributions, and predicted masses for $[8_c]_{c\bar{c}}\otimes [8_c]_{c\bar{c}}$ tetraquark satates .}
\label{tab1}
\end{table}

\begin{table}[htb]
\begin{center}
\begin{tabular}{|c|c|c|c|c|}\hline\hline
  & case A   & case B  & case C & case D   \\ \hline
 $[8_c]_{c\bar{c}}\otimes [8_c]_{b\bar{b}}$ & $12.51 \pm 0.10$      & $12.58 \pm 0.10$ & $12.67 \pm 0.10$ & $12.74 \pm 0.11$       \\ \hline
 $[8_c]_{c\bar{b}}\otimes [8_c]_{b\bar{c}}$ & $12.49 \pm 0.11$     & $12.58 \pm 0.10$ & $12.75 \pm 0.11$ & $12.81 \pm 0.10$   \\ \hline
 $[8_c]_{b\bar{b}}\otimes [8_c]_{b\bar{b}}$ & $18.38 \pm 0.11$      & $18.44 \pm 0.10$ & $18.50 \pm 0.10$ & $18.59 \pm 0.11$     \\
 \hline \hline
\end{tabular}
\end{center}
\caption{Predicted masses of the ground states for $[8_c]_{c\bar{c}}\otimes [8_c]_{b\bar{b}}$, $[8_c]_{c\bar{b}}\otimes [8_c]_{b\bar{c}}$, and $[8_c]_{b\bar{b}}\otimes [8_c]_{b\bar{b}}$, where cases A to D correspond to the currents in Eqs.(\ref{current-1}-\ref{current-4}).}
\label{tab2}
\end{table}

For $0^{++}$ fully-charm tetraquark states, we plot the mass curves as functions of the Borel parameter $M_B^2$ for different $\sqrt{s_0}$ in Figs.\ref{fig1}, where $A$ to $D$ stand for cases $A$ to $D$, and the two vertical lines in each case indicate the upper and lower bounds of the valid Borel window with the central value of $\sqrt{s_0}$. We search for the Borel parameter $M_B^2$ and continuum threshold parameters $s_0$ to satisfy the two criteria of the QCD sum rules: pole dominance at the phenomenological side and convergence of the operator product expansion at the QCD side. Furthermore, we take the relation $\sqrt{s_0} = M_X + (0.4-0.8) \text{GeV}$ as an additional constraint to obey.

The resulting Borel parameters, continuum threshold parameters, pole contributions are shown explicitly in Table.\ref{tab1}. From the table, we can see that the pole dominance at the phenomenological side is well satisfied. In the Borel windows, the contributions of the two-gluon condensate are smaller than the total contribution. Moreover, the mass curves in these Borel windows have the optimal platform. Now the three criteria of the QCD sum rules are all satisfied, we expect to make reasonable predictions.

After the above evaluation, we can then determine the masses of the scalar fully-charm tetraquark states with currents $A$ to $D$, that is
\begin{eqnarray}
M^{A,0^{++}}_X &=& (6.44 \pm 0.11) \, \text{GeV} \; , \\
M^{B,0^{++}}_X &=& (6.52 \pm 0.11) \, \text{GeV} \; , \\
M^{C,0^{++}}_X &=& (6.87 \pm 0.10) \, \text{GeV} \; , \\
M^{D,0^{++}}_X &=& (6.96 \pm 0.11) \, \text{GeV} \; , \label{eq-mass-1}
\end{eqnarray}
where the central values correspond to the results with the optimal stability of $M_B^2$, and the errors stem from the uncertainties of the condensates, the quark mass, the threshold parameter $\sqrt{s_0}$, and the Borel parameter $M_B^2$.

By swapping the flavor of the heavy quarks in the interpolating currents and performing the same calculation and analysis, we can obtain the corresponding results of the fully-heavy tetraquark states for $[8_c]_{c\bar{c}}\otimes [8_c]_{b\bar{b}}$, $[8_c]_{c\bar{b}}\otimes [8_c]_{b\bar{c}}$, and $[8_c]_{b\bar{b}}\otimes [8_c]_{b\bar{b}}$, whose masses are shown in Table.\ref{tab2}, respectively.

\section{Conclusions}
Very recently, the LHCb Collaboration reported their discovery of resonance-like structure in the di-$J/\Psi$ mass spectrum, which opens a whole new arena for multiquark exotic hadron. The narrow structure X(6900) is the first clear candidate for a multiquark exotic state that are composed of four charm quarks $cc \bar{c} \bar{c}$. This fully-heavy sector is particularly interesting from a theoretical point of view, since the molecular structure popular for multiquark states that contain light flavor is much less viable, leaving the color binding structures (such as the diquark-antidiquark structure and the octet-octet tetraquark structure) as the leading candidates.

In this work, we study the $J^{PC}=0^{++}$ fully-heavy tetraquark states via the QCD sum rules by constructing the octet-octet type currents. In the calculation, we consider the nonperturbative condensate contributions up to dimension 4 in the operator product expansion. Using the interpolating currents in Eqs.(\ref{current-1}-\ref{current-4}), we investigate all these four currents and collect the masses of these $[8_c]_{c\bar{c}}\otimes [8_c]_{c\bar{c}}$ in Table.\ref{tab1}. While comparing numerically to the recent LHCb results, we find the masses of currents (\ref{current-1}) and (\ref{current-2}) are apparently lower than X(6900), but in the scope of the broad structure from 6.2 to 6.8 GeV, this implies that these two configurations should make contributions to the broad structure as well as other possible configurations, for example, the $J/\psi-J/\psi$ molecule \cite{Albuquerque:2020hio}. Therefore the mixing effect is the reason that the structure in the range (6.2, 6.8) GeV observed by the LHCb Collaboration is a broad one. Particularly, since the sum rules of the currents (\ref{current-3}) and (\ref{current-4}) can really produce the mass of X(6900), we may probably conclude that the best interpretation is assigning X(6900) to the $0^{++}$ octet-octet tetraquark states with the configurations as currents (\ref{current-3}) and (\ref{current-4}).

Extending to the b-quark sector,the masses of their fully-bottom partners are found to lie in the region 18.38-18.59 GeV. Additionally, we also analyze the spectra of the $[8_c]_{c\bar{c}}\otimes [8_c]_{b\bar{b}}$ and $[8_c]_{c\bar{b}}\otimes [8_c]_{b\bar{c}}$ tetraquark states, which lie in the range of 12.51-12.74 GeV and 12.49-12.81 GeV, respectively.

\vspace{.7cm} {\bf Acknowledgments} \vspace{.3cm}

This work was supported in part by the National Natural Science Foundation of China(NSFC) under the Grants 11975236 and 11635009; Science Foundation of Hebei Normal University under Contract No. L2016B08; and National Key Research and Development Program of China under Contracts No. 2020YFA0406400..


\begin{widetext}


\appendix

\textbf{Appendix}

We list in this appendix the explicit expressions of the QCD spectral densities $\rho(s)$ and $\Pi(M_B^2)$ in Eq.(\ref{Pi-MB}) for all currents shown in Eqs. (\ref{current-1}- \ref{current-4}).

For the $[Q \bar{Q}] [Q \bar{Q}]$ octet-octet tetraquark states with $Q=c, b$, the spectral densities $\rho(s)$ and $\Pi(M_B^2)$ are
\begin{eqnarray}
\rho_{j, pert}^{0^{++}}(s) &=& \frac{1}{2^{7} \times \pi^{5}} \int_{16m_Q^{2}}^{s_{0}} ds \int^{x_{f}}_{x_{i}} dx \int^{y_{f}}_{y_{i}} dy \int^{z_{f}}_{z_{i}} dz \bigg \{ \frac{{\cal F}^{4}_{xyz} xyz {\cal A}_{xyz}}{2} \nonumber \\
&-& \frac{{\cal F}^{3}_{xyz} m_Q^{2}[6 \hat{s} xyz {\cal A}_{xyz}+{\cal N}_{j} 2 z {\cal A}_{xyz}+{\cal N}_{j} 2 xy]}{3} \nonumber \\
&+& {\cal F}^{2}_{xyz} m_Q^{4}[\hat{s}^{2} xyz {\cal A}_{xyz}+{\cal N}_{j} \hat{s} z {\cal A}_{xyz}+ {\cal N}_{j} \hat{s} xy+1] \bigg \} , \\
\rho_{j,\langle GG \rangle}^{0^{++}}(s) &=& \frac{\langle g_s^2 GG \rangle}{2^{13} \times 3^{2} \times \pi^{5} x^{3} y^{3} z^{3} {\cal A}_{xyz}^{3}} \int_{16m_Q^{2}}^{s_{0}} ds \int^{x_{f}}_{x_{i}} dx \int^{y_{f}}_{y_{i}} dy \int^{z_{f}}_{z_{i}} dz  \nonumber \\
&\times& \bigg \{ - \frac{{\cal F}^{2}_{xyz} [18x^{4} z^{3} y^{4} {\cal A}_{xyz}^{3}+18x^{3} y^{3} z^{4} {\cal A}_{xyz}^{4}]}{2}+{\cal F}_{xyz} m_Q^{2} \big [ {\cal A}_{xyz}^{4} [48 x y^{3} z^{4} \nonumber \\ &\times& (y-{\cal N}_{j})
+ 48 x^{4} y z (y^{3}+z^{3})+{\cal N}_{j} 12 x^{2} y^{2} z^{4}-{\cal N}_{j} 48 x^{3} y z^{4}+18\hat{s} x^{3} y^{3} z^{4}] \nonumber \\
&+& {\cal A}_{xyz}^{3}(18 \hat{s} x^{4} y^{4} z^{3}-{\cal N}_{j}48 x^{4} y^{4} z+{\cal N}_{j}12 x^{3} y^{3} z^{3})+12 x^{4} y^{4} z^{2} {\cal A}_{xyz}^{2} \nonumber \\
&+& 48 x^{4} y^{4} z^{3} (z-{\cal N}_{j}) {\cal A}_{xyz} \big ]-m_Q^{4} \big [{\cal A}_{xyz}^{4}[{\cal N}_{j}16x^{3} z^{4}+3\hat{s}^{2} x^{3} y^{3} z^{4}+{\cal N}_{j}16x^{3} y^{3} z \nonumber \\
&+& {\cal N}_{j}16y^{3} z^{4}+6{\cal N}_{j} \hat{s} x^{2} y^{2} z^{4}-{\cal N}_{j} 24\hat{s} x^{3} y z^{4}+24\hat{s} x y^{3} z^{4} (2y-{\cal N}_{j}) \nonumber \\
&+& 48\hat{s} x^{4} y z (y^{3}+z^{3})]+{\cal A}_{xyz}^{3}[{\cal N}_{j}16x^{4} y^{4}+3\hat{s}^{2} x^{4} y^{4} z^{3}+{\cal N}_{j} 6\hat{s} x^{3} y^{3} z^{3} \nonumber \\
&+& 6x^{2} y^{2} z^{3}+{\cal N}_{j}16x^{4} y z^{3}-24x^{3} y z^{3}+{\cal N}_{j}8x y^{3} z^{3} (2y-{\cal N}_{j}3) \nonumber \\
&-& 24x^{3} y^{3} z-{\cal N}_{j} 24\hat{s} x^{4} y^{4} z]+{\cal A}_{xyz}^{2}[6x^{3} y^{3} z^{2}+{\cal N}_{j}6\hat{s} x^{4} y^{4} z^{2}] \nonumber \\
&+& {\cal A}_{xyz}[{\cal N}_{j}8x^{3} y{3} z^{3} (2z-{\cal N}_{j}3)+24\hat{s} x^{4} y^{4} z^{3} (2z-{\cal N}_{j})]+{\cal N}_{j}16x^{4} y^{4} z^{3} \big ] \bigg \}, \\
\Pi_{j,\langle GG \rangle}^{0^{++}}(M_B^2) &=& - \frac{\langle g_s^2 GG \rangle}{2^{9} \times 3^{2}\pi^{6}} \int^{1}_{0} dx \int^{1-x}_{0} dy \int^{1-x-y}_{0} dze^{- \frac{f_{xyz} m_Q^{2}}{M_B^2}} \nonumber \\
&\times& \frac{1}{ x^{3} y^{3} z^{3} {\cal A}_{xyz}^{3}} m_Q^{6}   (f_{xyz} z {\cal A}_{xyz}+{\cal N}_{j}) (f_{xyz} xy+{\cal N}_{j}) \nonumber \\ &\times& \bigg \{ {\cal A}_{xyz}^{3}[x^{3}(y^{3}+z^{3})+y^{3} z^{3}]+x^{3} y^{3} z^{3} \bigg \} ,
\end{eqnarray}

\begin{eqnarray}
\rho_{k, pert}^{0^{++}}(s) &=& \frac{1}{2^{6} \times \pi^{5}} \int_{16m_Q^{2}}^{s_{0}} ds \int^{x_{f}}_{x_{i}} dx \int^{y_{f}}_{y_{i}} dy \int^{z_{f}}_{z_{i}} dz \bigg \{ {\cal F}^{4}_{xyz} xyz {\cal A}_{xyz} \nonumber \\
&-& \frac{{\cal F}^{3}_{xyz} m_Q^{2}[12 \hat{s} xyz {\cal A}_{xyz}+{\cal N}_{k} 2 z {\cal A}_{xyz}+{\cal N}_{k} 2 xy]}{3} \nonumber \\
&+& {\cal F}^{2}_{xyz} m_Q^{4}[2\hat{s}^{2} xyz {\cal A}_{xyz}+{\cal N}_{k} \hat{s} z {\cal A}_{xyz}+ {\cal N}_{k} \hat{s} xy+2] \bigg \} , \\
\rho_{k,\langle GG \rangle}^{0^{++}}(s) &=& \frac{\langle g_s^2 GG \rangle m_Q^{2}}{2^{10} \times 3^{2} \times \pi^{5} } \int_{16m_Q^{2}}^{s_{0}} ds \int^{x_{f}}_{x_{i}} dx \int^{y_{f}}_{y_{i}} dy \int^{z_{f}}_{z_{i}} dz  \frac{1}{x^{3} y^{3} z^{3} {\cal A}_{xyz}^{3}} \nonumber \\
&\times& \bigg \{6{\cal F}_{xyz} xyz {\cal A}_{xyz} \big [{\cal A}_{xyz}^{3}[8x^{3}(y^{3}+z^{3})-{\cal N}_{k}4x^{2}z^{3}+4y^{2}z^{3}(2y-{\cal N}_{k})] \nonumber \\
&-& {\cal N}_{k}x^{2} y^{2} (4xy+z^{2}) {\cal A}_{xyz}^{2}+4x^{3} y^{3} z^{2} (2z-{\cal N}_{k}) \big ]-m_Q^{2} \big [4z {\cal A}_{xyz}^{4}[12\hat{s} x^{4} y \nonumber \\ &\times& (y^{3}+z^{3})
+ {\cal N}_{k}x^{3}(-3\hat{s} y z^{3}+2y^{3}+2z^{3})+3\hat{s} x y^{3} z^{3} (4y-{\cal N}_{k})+{\cal N}_{k}2y^{3} z^{3}] \nonumber \\
&-& xy{\cal A}_{xyz}^{3} [4x^{3}{\cal N}_{k}\big (y^{3}(3\hat{s}-2)-2z^{3} \big )+{\cal N}_{k}3x^{2} \big (y^{2}z(\hat{s} z^{2}+8{\cal N}_{k})+{\cal N}_{k}8z^{3} \big ) \nonumber \\
&-& {\cal N}_{k}8y^{2} z^{3} (y-3{\cal N}_{k})]+4x^{3} y^{3} z^{3} {\cal A}_{xyz} \big (3\hat{s} xy(4z-{\cal N}_{k})+2{\cal N}_{k}(z-3{\cal N}_{k}) \big ) \nonumber \\
&+& {\cal N}_{k}8x^{4} y^{4} z^{3} \big ] \bigg \} ,\\
\Pi_{k,\langle GG \rangle}^{0^{++}}(M_B^2) &=& -\frac{\langle g_s^2 GG \rangle}{2^{8} \times 3^{2} \times \pi^{6}} \int^{1}_{0} dx \int^{1-x}_{0} dy \int^{1-x-y}_{0} dz e^{- \frac{f_{xyz} m_Q^{2}}{M_B^2}} \nonumber \\
&\times& \frac{1}{ x^{3} y^{3} z^{3} {\cal A}_{xyz}^{3}} m_Q^{6} \{ {\cal A}_{xyz}^{3}[x^{3}(y^{3}+z^{3})+y^{3} z^{3}]+x^{3} y^{3} z^{3} \} \nonumber \\
&\times& [2 f^{2}_{xyz} xyz {\cal A}_{xyz}+{\cal N}_{k} f_{xyz}(z {\cal A}_{xyz}+xy)+2] ,
\end{eqnarray}
where the subscript $j$ represents $A$ and $C$, and $k$ denotes $B$ and $D$, and the factors ${\cal N}_j$ and ${\cal N}_k$ have the following definition: ${\cal N}_A ={\cal N}_B=1$ and ${\cal N}_C = {\cal N}_D = -1$.

For the $[c \bar{c}] [b \bar{b}]$ octet-octet tetraquark states, the spectral densities $\rho(s)$ and $\Pi(M_B^2)$ read
\begin{eqnarray}
\rho_{j, pert}^{0^{++}}(s) &=& \frac{1}{2^{8} \times \pi^{5}} \int_{(2m_{c}+2m_{b})^{2}}^{s_{0}} ds \int^{x_{f}}_{x_{i}} dx \int^{y_{f}}_{y_{i}} dy \int^{z_{f}}_{z_{i}} dz \bigg \{ \frac{{\cal F}^{4}_{xyz} xyz {\cal A}_{xyz}}{2} \nonumber \\
&-& \frac{{\cal F}^{3}_{xyz} [6m_{b}^{2} \hat{s} xyz {\cal A}_{xyz}+{\cal N}_{j} 2m_{c}^{2} z {\cal A}_{xyz}+{\cal N}_{j} 2m_{b}^{2} xy]}{3} \nonumber \\
&+& {\cal F}^{2}_{xyz} [m_{b}^{4} \hat{s}^{2} xyz {\cal A}_{xyz}+{\cal N}_{j} m_{b}^{2} m_{c}^{2} \hat{s} z {\cal A}_{xyz}+ {\cal N}_{j} m_{b}^{4} \hat{s} xy+m_{b}^{2} m_{c}^{2}] \bigg \} , \\
\rho_{j,\langle GG \rangle}^{0^{++}}(s) &=& \frac{\langle g_s^2 GG \rangle}{2^{13} \times 3^{2} \times \pi^{5}} \int_{(2m_{c}+2m_{b})^{2}}^{s_{0}} ds \int^{x_{f}}_{x_{i}} dx \int^{y_{f}}_{y_{i}} dy \int^{z_{f}}_{z_{i}} dz \frac{1}{x^{3} y^{3} z^{3} {\cal A}_{xyz}^{3}} \nonumber \\
&\times& \bigg \{ - \frac{{\cal F}^{2}_{xyz} [18x^{3} z^{3} y^{4} {\cal A}_{xyz}^{4}+18x^{4} y^{4} z^{3} {\cal A}_{xyz}^{3}]}{2}+{\cal F}_{xyz} \big [{\cal A}_{xyz}^{4}[48 x^{4} y z (m_{b}^{2} y^{3}+m_{c}^{2} z^{3}) \nonumber \\
&+& 18 m_{b}^{2} \hat{s} x^{3} y^{3} z^{4}-{\cal N}_{j} 48m_{c}^{2} x^{3} y z^{4}+{\cal N}_{j} 12m_{c}^{2} x^{2} y^{2} z^{4}+48m_{c}^{2} x y^{3} z^{4} (y-{\cal N}_{j})] \nonumber \\
&+& {\cal A}_{xyz}^{3}[18 m_{b}^{2} \hat{s} x^{4} y^{4} z^{3}-{\cal N}_{j} 48m_{b}^{2} x^{4} y^{4} z+{\cal N}_{j} 6m_{b}^{2} x^{3} y^{3} z^{3}+{\cal N}_{j} 6m_{c}^{2} x^{3} y^{3} z^{3}] \nonumber \\
&+& {\cal A}_{xyz}^{2} {\cal N}_{j}12 m_{b}^{2} x^{4} y^{4} z^{2}+48 m_{b}^{2} x^{4} y^{4} z^{3} (z-{\cal N}_{j}) {\cal A}_{xyz}\big ]-\big [{\cal A}_{xyz}^{4}[3 m_{b}^{4} \hat{s}^{2} x^{3} y^{3} z^{4} \nonumber \\
&+& 48 m_{b}^{2} \hat{s} x^{4} y z (m_{b}^{2} y^{3}+m_{c}^{2} z^{3})-{\cal N}_{j}24 m_{b}^{2} m_{c}^{2} \hat{s} x^{3} y z^{4}+{\cal N}_{j} 6 m_{b}^{2} m_{c}^{2} \hat{s} x^{2} y^{2} z^{4} \nonumber \\
&+& 24 m_{b}^{2} m_{c}^{2} \hat{s} x y^{3} z^{4} (2y-{\cal N}_{j})+{\cal N}_{j} 16\hat{s} x y^{3} z^{4} (2y-{\cal N}_{j}) \nonumber \\
&+& {\cal N}_{j}16m_{b}^{2} m_{c}^{2} x^{3} y^{3} z]+{\cal A}_{xyz}^{3}[3m_{b}^{4} \hat{s}^{2} x^{4} y^{4} z^{3}+6m_{b}^{2} m_{c}^{2} x^{2} y^{2} z^{3}-24 m_{b}^{2} m_{c}^{2} x^{3} y z^{3} \nonumber \\
&-& {\cal N}_{j}24m_{b}^{4} \hat{s} x^{4} y^{4} z+{\cal N}_{j}16m_{b}^{4} x^{4} y^{4}+{\cal N}_{j}3m_{b}^{2}(m_{b}^{2}+m_{c}^{2}) \hat{s} x^{3} y^{3} z^{3} \nonumber \\
&+& {\cal N}_{j}16m_{b}^{2} m_{c}^{2} x^{4} y z^{3}-24 m_{b}^{2} m_{c}^{2} x^{3} y^{3} z+{\cal N}_{j} 8m_{b}^{2} m_{c}^{2} x y^{3} z^{3} (2y-3{\cal N}_{j})] \nonumber \\
&+& {\cal A}_{xyz}^{2}[{\cal N}_{j}6m_{b}^{4} \hat{s} x^{4} y^{4} z^{2}+6m_{b}^{2} m_{c}^{2} x^{3} y{3} z^{2}]+{\cal A}_{xyz}[24m_{b}^{4} \hat{s} x^{4} y^{4} z^{3} (2z-{\cal N}_{j}) \nonumber \\
&+& {\cal N}_{j}8m_{b}^{2} m_{c}^{2} x^{3} y^{3} z^{3} (2z-3{\cal N}_{j})]+{\cal N}_{j}16m_{b}^{4} x^{4} y^{4} z^{3} \nonumber \\
&+& {\cal N}_{j}16m_{c}^{4} x^{3} z^{4}+{\cal N}_{j}16m_{c}^{4} y^{3} z^{4} \big ] \bigg \}, \\
\Pi_{j,\langle GG \rangle}^{0^{++}}(M_B^2) &=& - \frac{\langle g_s^2 GG \rangle}{2^{10} \times 3^{2} \times \pi^{6}} \int^{1}_{0} dx \int^{1-x}_{0} dy \int^{1-x-y}_{0} dz e^{- \frac{f_{xyz} m_b^{2}}{M_B^2}}\nonumber \\
&\times& \frac{1}{ x^{3} y^{3} z^{3} {\cal A}_{xyz}^{3}} m_b^{2} (f_{xyz} z {\cal A}_{xyz}+{\cal N}_{j}) (f_{xyz} m_{b}^{2} xy+{\cal N}_{j}m_{c}^{2}) \nonumber \\
&\times& [m_{b}^{2} x^{3} y^{3}({\cal A}_{xyz}^{3}+z^{3})+{\cal A}_{xyz}^{3} m_{c}^{2} z^{3}(x^{3}+y^{3})] ,
\end{eqnarray}

\begin{eqnarray}
\rho_{k, pert}^{0^{++}}(s) &=& \frac{1}{2^{7} \times \pi^{5}} \int_{(2m_{c}+2m_{b})^{2}}^{s_{0}} ds \int^{x_{f}}_{x_{i}} dx \int^{y_{f}}_{y_{i}} dy \int^{z_{f}}_{z_{i}} dz \bigg \{ {\cal F}^{4}_{xyz} xyz {\cal A}_{xyz} \nonumber \\
&-& \frac{{\cal F}^{3}_{xyz} [12m_{b}^{2} \hat{s} xyz {\cal A}_{xyz}+{\cal N}_{k} 2m_{c}^{2} z {\cal A}_{xyz}+{\cal N}_{k} 2m_{b}^{2} xy]}{3} \nonumber \\
&+& {\cal F}^{2}_{xyz} [2m_{b}^{4} \hat{s}^{2} xyz {\cal A}_{xyz}+{\cal N}_{k} m_{b}^{2} m_{c}^{2} \hat{s} z {\cal A}_{xyz}+ {\cal N}_{k} m_{b}^{4} \hat{s} xy+m_{b}^{2} m_{c}^{2}] \bigg \} , \\
\rho_{k,\langle GG \rangle}^{0^{++}}(s) &=& \frac{\langle g_s^2 GG \rangle}{2^{12} \times 3^{2}\pi^{5}} \int_{(2m_{c}+2m_{b})^{2}}^{s_{0}} ds \int^{x_{f}}_{x_{i}} dx \int^{y_{f}}_{y_{i}} dy \int^{z_{f}}_{z_{i}} dz
\frac{1}{x^{3} y^{3} z^{3} {\cal A}_{xyz}^{3}} \nonumber \\ &\times& \bigg \{ {\cal F}_{xyz} \big [{\cal A}_{xyz}^{4}[96m_{b}^2 x^{4} y^{4} z+96m_{c}^2 x^{4} y z^{4}-{\cal N}_{k} 48m_{c}^{2} x^{3} y z^{4} \nonumber \\
&+& 48m_{c}^{2} xy^{3} z^{4} (2y-{\cal N}_{k})]-{\cal A}_{xyz}^{3}({\cal N}_{k} 48m_{b}^{2} x^{4} y^{4} z-{\cal N}_{k}6m_{b}^{2} x^{3} y^{3} z^{3} \nonumber \\
&-& {\cal N}_{k}6m_{cb}^{2} x^{3} y^{3} z^{3})+48m_{b}^{2} x^{4} y^{4} z^{3} (2z-{\cal N}_{k}) {\cal A}_{xyz} \big ]-\big [{\cal A}_{xyz}^{4}[96m_{b}^{4} \hat{s} x^{4} y^{4} z  \nonumber \\
&+& 96m_{b}^{2} m_{c}^{2} \hat{s} x^{4} y z^{4}-{\cal N}_{k}24m_{b}^{2} m_{c}^{2} \hat{s} x^{3} y z^{4}+24m_{b}^{2} m_{c}^{2} \hat{s} x y^{3} z^{4} (4y-{\cal N}_{k}) \nonumber \\
&+& {\cal N}_{k}16m_{b}^{2} m_{c}^{2} x^{3} y^{3} z+{\cal N}_{k}16m_{c}^{4} x^{3} z^{4}+{\cal N}_{k}16m_{c}^{4} y^{3} z^{4}] \nonumber \\
&-& {\cal A}_{xyz}^{3}({\cal N}_{k}24m_{b}^{4} \hat{s} x^{4} y^{4} z-{\cal N}_{k}3m_{b}^{4} \hat{s} x^{3} y^{3} z^{3})+{\cal N}_{k}16m_{b}^{4} x^{4} y^{4} ({\cal A}_{xyz}^{3}+z^{3}) \nonumber \\
&-& {\cal A}_{xyz}^{3}[{\cal N}_{k}3m_{b}^{2} m_{c}^{2} \hat{s} x^{3} y^{3} z^{3}+{\cal N}_{k}16m_{b}^{2} m_{c}^{2} x^{4} y z^{3}-48m_{b}^{2} m_{c}^{2} x^{3} y^{3} z \nonumber \\
&-& 48m_{b}^{2} m_{c}^{2} x^{3} y z^{3}+{\cal N}_{k}16m_{b}^{2} m_{c}^{2} x y^{3} z^{3} (y-3{\cal N}_{k})] \nonumber \\
&+& {\cal A}_{xyz}[24m_{b}^{4}\hat{s} x^{4} y^{4} z^{3} (4z-{\cal N}_{k})+{\cal N}_{k}16m_{b}^{2} m_{c}^{2} x^{3} y^{3} z^{3} (z-3{\cal N}_{k})] \big ] \bigg \} ,\\
\Pi_{k,\langle GG \rangle}^{0^{++}}(M_B^2) &=& -\frac{\langle g_s^2 GG \rangle}{2^{9} \times 3^{2} \times \pi^{6}} \int^{1}_{0} dx \int^{1-x}_{0} dy \int^{1-x-y}_{0} dz e^{- \frac{f_{xyz} m_b^{2}}{M_B^2}} \nonumber \\
&\times& \frac{m_b^{2}}{ x^{3} y^{3} z^{3} {\cal A}_{xyz}^{3}}  [f_{xyz} m_{b}^{2} xy(2{\cal A}_{xyz} f_{xyz}z+{\cal N}_{k})
+ m_{c}^{2}({\cal N}_{k} {\cal A}_{xyz} f_{xyz}z+2)]\nonumber \\ &\times& [m_{b}^{2} x^{3} y^{3}({\cal A}_{xyz}^{3}+z^{3})+{\cal A}_{xyz}^{3} m_{c}^{2} z^{3}(x^{3}+y^{3})].
\end{eqnarray}

For the $[c \bar{b}] [b \bar{c}]$ octet-octet tetraquark states, the spectral densities $\rho(s)$ and $\Pi(M_B^2)$ can be written as
\begin{eqnarray}
\rho_{j, pert}^{0^{++}}(s) &=& \frac{1}{2^{7} \times \pi^{5}} \int_{(2m_{c}+2m_{b})^{2}}^{s_{0}} ds \int^{x_{f}}_{x_{i}} dx \int^{y_{f}}_{y_{i}} dy \int^{z_{f}}_{z_{i}} dz \bigg \{ \frac{{\cal F}^{4}_{xyz} xyz {\cal A}_{xyz}}{2} \nonumber \\
&-& \frac{{\cal F}^{3}_{xyz} (6m_{b}^{2} \hat{s} xyz {\cal A}_{xyz}+{\cal N}_{j} 2m_{b} m_{c} x {\cal A}_{xyz}+{\cal N}_{j} 2m_{b} m_{c} yz)}{3} \nonumber \\
&+& {\cal F}^{2}_{xyz} [m_{b}^{4} \hat{s}^{2} xyz {\cal A}_{xyz}+{\cal N}_{j} m_{b}^{3} m_{c} \hat{s} x {\cal A}_{xyz}+ {\cal N}_{j} m_{b}^{3} m_{c} \hat{s} yz+m_{b}^{2} m_{c}^{2}] \bigg \} , \\
\rho_{j,\langle GG \rangle}^{0^{++}}(s) &=& \frac{\langle g_s^2 GG \rangle}{2^{12} \times 3^{2} \times \pi^{5} } \int_{(2m_{c}+2m_{b})^{2}}^{s_{0}} ds \int^{x_{f}}_{x_{i}} dx \int^{y_{f}}_{y_{i}} dy \int^{z_{f}}_{z_{i}} dz \frac{1}{x^{3} y^{3} z^{3} {\cal A}_{xyz}^{3}}  \nonumber \\
&\times& \bigg \{ - \frac{{\cal F}^{2}_{xyz} [18x^{4} z^{3} y^{3} {\cal A}_{xyz}^{4}+18x^{3} y^{4} z^{4} {\cal A}_{xyz}^{3}]}{2}+{\cal F}_{xyz} \big [{\cal A}_{xyz}^{4}[18m_{b}^{2} \hat{s} x^{4} y^{3} z^{3} \nonumber \\
&+& 48 m_{b}^{2} x^{4} y^{4} z-{\cal N}_{j} 48m_{b} m_{c} x^{4} y^{3} z+{\cal N}_{j} 12m_{b} m_{c} x^{4} y^{2} z^{2} \nonumber \\
&+& 48m_{c} x^{4} y z^{3} (m_{c}z-{\cal N}_{j} m_{b})+48m_{c}^{2}x y^{4} z^{4}]+{\cal A}_{xyz}^{3}[18m_{b}^{2} \hat{s} x^{3} y^{4} z^{4} \nonumber \\
&+& {\cal N}_{j} 12m_{b} m_{c} x^{3} y^{3} z^{3}-{\cal N}_{j} 48m_{b} m_{c} x y^{4} z^{4}]+{\cal N}_{j} 12m_{b} m_{c} x^{2} y^{4} z^{4} {\cal A}_{xyz}^{2} \nonumber \\
&+& 48m_{b}^{2} x^{4} y^{4} z^{4} {\cal A}_{xyz}-{\cal N}_{j} 48m_{b} m_{c} x^{3} y^{4} z^{4} \big ]-\big [{\cal A}_{xyz}^{4}[3m_{b}^{4} \hat{s}^{2} x^{4} y^{3} z^{3} \nonumber \\
&+& 48m_{b}^{4} \hat{s} x^{4} y^{4} z-{\cal N}_{j} 24m_{b}^{3} m_{c} \hat{s} x^{4} y^{3} z+{\cal N}_{j} 6m_{b}^{3} m_{c} \hat{s} x^{4} y^{2} z^{2} \nonumber \\
&+& {\cal N}_{j} 16m_{b}^{3} m_{c} x^{4} y^{3}+48m_{b}^{2} m_{c}^{2} \hat{s} x y^{4} z^{4}-{\cal N}_{j} 24m_{b}^{2} m_{c} \hat{s} x^{4} y z^{3} (m_{b}-{\cal N}_{j}2m_{c} z) \nonumber \\
&+& {\cal N}_{j} 16m_{b} m_{c}^{3} x^{4} z^{3}+{\cal N}_{j} 16m_{b} m_{c}^{3} x y^{3} z^{3}]+{\cal A}_{xyz}^{3}[3m_{b}^{4} \hat{s}^{2} x^{3} y^{4} z^{4} \nonumber \\
&+& {\cal N}_{j} 6m_{b}^{3} m_{c} \hat{s} x^{3} y^{3} z^{3}-{\cal N}_{j} 24m_{b}^{3} m_{c} \hat{s} x y^{4} z^{4} +{\cal N}_{j} 16m_{b}^{3} m_{c} \hat{s} x^{3} y^{4} z \nonumber \\
&-& 24m_{b}^{2} m_{c}^{2} x^{3} y^{3} z+6m_{b}^{2} m_{c}^{2} x^{3} y^{2} z^{2}-24m_{b}^{2} m_{c}^{2} x y^{3} z^{3}+{\cal N}_{j} 16m_{b} m_{c}^{3} y^{4} z^{4} \nonumber \\
&+& {\cal N}_{j} 8m_{b} m_{c}^{2} x^{3} y z^{3} (2m_{c}z-{\cal N}_{j} 3m_{b})]+{\cal A}_{xyz}^{2}[{\cal N}_{j} 6m_{b}^{3} m_{c} \hat{s} x^{2} y^{4} z^{4} \nonumber \\
&+& 6m_{b}^{2} m_{c}^{2} x^{2} y^{3} z^{3}]+{\cal A}_{xyz}[48m_{b}^{4} \hat{s} x^{4} y^{4} z^{4}-{\cal N}_{j} 24m_{b}^{3} m_{c} \hat{s} x^{3} y^{4} z^{4} \nonumber \\
&+& {\cal N}_{j} 16m_{b}^{3} m_{c} x^{4} y^{3} z^{3}-24m_{b}^{2} m_{c}^{2} x^{3} y^{3} z^{3}]+{\cal N}_{j} 16m_{b}^{3} m_{c} x^{3} y^{4} z^{4} \big ] \bigg \} ,\\
\Pi_{j,\langle GG \rangle}^{0^{++}}(M_B^2) &=& - \frac{\langle g_s^2 GG \rangle}{2^{9} \times 3^{2} \times \pi^{6}} \int^{1}_{0} dx \int^{1-x}_{0} dy \int^{1-x-y}_{0} dz e^{- \frac{f_{xyz} m_b^{2}}{M_B^2}} \nonumber \\
&\times& \frac{1}{x^{3} y^{3} z^{3} {\cal A}_{xyz}^{3}} m_b^{2} ({\cal N}_{j} f_{xyz} m_{b} x {\cal A}_{xyz}+m_{c})   ({\cal N}_{j} f_{xyz} m_{b} yz+m_{c}) \nonumber \\
&\times& [m_{b}^{2} x^{3} y^{3}({\cal A}_{xyz}^{3}+z^{3})+{\cal A}_{xyz}^{3} m_{c}^{2} z^{3}(x^{3}+y^{3})] ,
\end{eqnarray}

\begin{eqnarray}
\rho_{k, pert}^{0^{++}}(s) &=& \frac{1}{2^{6} \times \pi^{5}} \int_{(2m_{c}+2m_{b})^{2}}^{s_{0}} ds \int^{x_{f}}_{x_{i}} dx \int^{y_{f}}_{y_{i}} dy \int^{z_{f}}_{z_{i}} dz \bigg \{ {\cal F}^{4}_{xyz} xyz {\cal A}_{xyz} \nonumber \\
&-& \frac{{\cal F}^{3}_{xyz} [12m_{b}^{2} \hat{s} xyz {\cal A}_{xyz}+{\cal N}_{k} 2m_{b} m_{c} x {\cal A}_{xyz}+{\cal N}_{k} 2m_{b} m_{c} yz]}{3} \nonumber \\
&+& {\cal F}^{2}_{xyz} [2m_{b}^{4} \hat{s}^{2} xyz {\cal A}_{xyz}+{\cal N}_{k} m_{b}^{3} m_{c} \hat{s} x {\cal A}_{xyz}+ {\cal N}_{k} m_{b}^{3} m_{c} \hat{s} yz+2m_{b}^{2} m_{c}^{2}] \bigg \} , \\
\rho_{k,\langle GG \rangle}^{0^{++}}(s) &=& \frac{\langle g_s^2 GG \rangle}{2^{10} \times 3^{2} \times \pi^{5}} \int_{(2m_{c}+2m_{b})^{2}}^{s_{0}} ds \int^{x_{f}}_{x_{i}} dx \int^{y_{f}}_{y_{i}} dy \int^{z_{f}}_{z_{i}} dz \frac{1}{x^{3} y^{3} z^{3} {\cal A}_{xyz}^{3}} \nonumber \\
&\times& \bigg \{ {\cal F}_{xyz} \big [{\cal A}_{xyz}^{4}[48m_{b}^2 x^{4} y^{4} z-{\cal N}_{k}24m_{b} m_{c} x^{4} y^{3} z-{\cal N}_{k}24m_{b} m_{c} x^{4} y z^{3} \nonumber \\
&+& 48m_{c}^{2} xy z^{4}(x^{3}+y^{3})]-{\cal A}_{xyz}^{3}[{\cal N}_{k} 6m_{b} m_{c} x^{3} y^{3} z^{3}+{\cal N}_{k} 24m_{b} m_{c} x y^{4} z^{4}] \nonumber \\
&+& {\cal A}_{xyz}[48m_{b}^{2} x^{4} y^{4} z^{4}-{\cal N}_{k} 24m_{b} m_{c} x^{3} y^{4} z^{4}] \big ]-\big [{\cal A}_{xyz}^{4}[-{\cal N}_{k} 12m_{b}^{3} m_{c} \hat{s} x^{4} y^{3} z  \nonumber \\
&-& {\cal N}_{k} 12m_{b}^{3} m_{c} \hat{s} x^{4} y z^{3}+48m_{b}^{2} m_{c}^{2} \hat{s} x y z^{4} (x^{3}+y^{3})+{\cal N}_{k} 8m_{b} m_{c}^{3} x^{4} z^{3} \nonumber \\
&+& {\cal N}_{k} 8m_{b} m_{c}^{3} x y^{3} z^{3}]+48m_{b}^{4} \hat{s} x^{4} y^{4} z ({\cal A}_{xyz}^{3}+z^{3}) {\cal A}_{xyz}-{\cal A}_{xyz}^{3}[{\cal N}_{k} 3m_{b}^{3} m_{c} \hat{s} \nonumber \\
&\times& x^{3} y^{3} z^{3} + {\cal N}_{k} 12m_{b}^{3} m_{c} \hat{s} x y^{4} z^{4}+24m_{b}^{2} m_{c}^{2} x^{3} y^{3} z+24m_{b}^{2} m_{c}^{2} x^{3} y z^{3} \nonumber \\
&+& 24m_{b}^{2} m_{c}^{2} x y^{3} z^{3}-{\cal N}_{k} 8m_{b} m_{c}^{3} x^{3} y z^{4}-{\cal N}_{k} 8m_{b} m_{c}^{3} y^{4} z^{4}] \nonumber \\
&-& {\cal A}_{xyz}[{\cal N}_{k} 12m_{b}^{3} m_{c} \hat{s} x^{3} y^{4} z^{4}+24m_{b}^{2} m_{c}^{2} x^{3} y^{3} z^{3}] \nonumber \\
&+& {\cal N}_{k} 8m_{b}^{3} m_{c} x^{3} y^{3} ({\cal A}_{xyz}^{3}+z^{3})({\cal A}_{xyz}x+yz) \big ] \bigg \} ,\\
\Pi_{k,\langle GG \rangle}^{0^{++}}(M_B^2) &=& - \frac{\langle g_s^2 GG \rangle}{2^{8} \times 3^{2} \times \pi^{6}} \int^{1}_{0} dx \int^{1-x}_{0} dy \int^{1-x-y}_{0} dz e^{- \frac{f_{xyz} m_b^{2}}{M_B^2}} \nonumber \\
&\times& \frac{m_b^{2}}{ x^{3} y^{3} z^{3} {\cal A}_{xyz}^{3}}  [2f_{xyz}^{2} m_{b}^{2} xyz {\cal A}_{xyz}+2m_{c}^{2}
+ {\cal N}_{k} f_{xyz} m_{b} m_{c} ({\cal A}_{xyz}x+yz)]
\nonumber \\ &\times& [m_{b}^{2} x^{3} y^{3}({\cal A}_{xyz}^{3}+z^{3})+{\cal A}_{xyz}^{3} m_{c}^{2} z^{3}(x^{3}+y^{3})].
\end{eqnarray}

Here, we also have the following definitions:
\begin{eqnarray}
  {\cal A}_{x} &=& (1-x), \;\;
  {\cal A}_{xy} = (1 - x - y), \;\;
  {\cal A}_{xyz} = (1 - x - y - z), \\
  f_{xyz} &=& \left[ \frac{r_m^2}{x} + \frac{r_m^2}{y} +\frac{1}{z} + \frac{1}{(1 - x - y - z)} \right], \\
  x_{f/i} &=& \frac{\hat{s} - (4 r_m + 1) \pm \sqrt{(\hat{s} -4)(\hat{s}- 4(r_m + 1)^2)}}{2 \hat{s}},  \\
  y_{f/i} &=& \frac{1}{2(\hat{s} x-r_m^2)}\bigg\{x(\hat{s} {\cal A}_{x} + r_m^2 - 4) - r_m^2 {\cal A}_{x} \nonumber \\
  &\pm& \sqrt{4 r_m^2 x {\cal A}_{x} (r_m^2 - \hat{s} x) + [r_m^2 {\cal A}_{x} - x (\hat{s} {\cal A}_{x} + r_m^2 -4)]}\bigg\},  \\
  z_{f/i} &=& \frac{1}{2[\hat{s} x y - r_m^2 (x+y)]}\bigg\{{\cal A}_{xy} [\hat{s} x y - r_m^2 (x +y)] \nonumber \\
  &\pm& \sqrt{{\cal A}_{xy}[{\cal A}_{xy} (\hat{s} x y - r_m^2 (x + y))^2 - 4 x y (\hat{s} x y -r_m^2 (x +y))]}\bigg\}.
\end{eqnarray}

It should be noted that for $[Q \bar{Q}][Q \bar{Q}]$ case with $Q =c$ or $b$, we have the following definitions: $\hat{s} = \frac{s}{m_Q^2}$, ${\cal F}_{xyz} = m_Q^2 f_{xyz} - s$, and $r_m =1$; for $[c \bar{c}][b \bar{b}]$ and $[c \bar{b}][b \bar{c}]$ cases, $\hat{s} = \frac{s}{m_b^2}$, ${\cal F}_{xyz} = m_b^2 f_{xyz} - s$, and $r_m =m_c/m_b$.

\end{widetext}


\begin{thebibliography}{9}

\bibitem{GellMann:1964nj}
  M.~Gell-Mann,
  Phys.\ Lett.\  {\bf 8}, 214 (1964).

\bibitem{Zweig}
  G.~Zweig, Report No. CERN-TH-401.

\bibitem{Choi:2003ue}
  S.~K.~Choi \textit{et al.} [Belle],
  Phys. Rev. Lett. \textbf{91}, 262001 (2003)
  [arXiv:hep-ex/0309032 [hep-ex]].

\bibitem{Aaij:2020fnh}
  R.~Aaij \textit{et al.} [LHCb],
  Sci. Bull. \textbf{2020}, 65
  [arXiv:2006.16957 [hep-ex]].

\bibitem{Khachatryan:2016ydm}
  V.~Khachatryan \textit{et al.} [CMS],
  JHEP \textbf{05}, 013 (2017)
  [arXiv:1610.07095 [hep-ex]].

\bibitem{Aaij:2018zrb}
  R.~Aaij \textit{et al.} [LHCb],
  JHEP \textbf{10}, 086 (2018)
  [arXiv:1806.09707 [hep-ex]].

\bibitem{Sirunyan:2020txn}
  A.~M.~Sirunyan \textit{et al.} [CMS],
  Phys. Lett. B \textbf{808}, 135578 (2020)
  [arXiv:2002.06393 [hep-ex]].

\bibitem{Iwasaki:1976cn}
  Y.~Iwasaki,
  Phys. Rev. Lett. \textbf{36}, 1266 (1976).

\bibitem{Chao:1980dv}
  K.~T.~Chao,
  Z. Phys. C \textbf{7}, 317 (1981).

\bibitem{Ader:1981db}
  J.~P.~Ader, J.~M.~Richard and P.~Taxil,
  Phys. Rev. D \textbf{25}, 2370 (1982).

\bibitem{Badalian:1985es}
  A.~M.~Badalian, B.~L.~Ioffe and A.~V.~Smilga,
  Nucl. Phys. B \textbf{281}, 85 (1987).

\bibitem{Heller:1985cb}
  L.~Heller and J.~A.~Tjon,
  Phys. Rev. D \textbf{32}, 755 (1985).


\bibitem{Anwar:2017toa}
  M.~N.~Anwar, J.~Ferretti, F.~K.~Guo, E.~Santopinto and B.~S.~Zou,
  Eur. Phys. J. C \textbf{78}, no.8, 647 (2018)
  [arXiv:1710.02540 [hep-ph]].

\bibitem{Bedolla:2019zwg}
  M.~A.~Bedolla, J.~Ferretti, C.~D.~Roberts and E.~Santopinto,
  Eur. Phys. J. C \textbf{80}, no.11, 1004 (2020).

\bibitem{Lloyd:2003yc}
  R.~J.~Lloyd and J.~P.~Vary,
  Phys. Rev. D \textbf{70}, 014009 (2004)
  [arXiv:hep-ph/0311179 [hep-ph]].

\bibitem{Barnea:2006sd}
  N.~Barnea, J.~Vijande and A.~Valcarce,
  Phys. Rev. D \textbf{73}, 054004 (2006)
  [arXiv:hep-ph/0604010 [hep-ph]].

\bibitem{Debastiani:2017msn}
  V.~R.~Debastiani and F.~S.~Navarra,
  Chin. Phys. C \textbf{43}, no.1, 013105 (2019)
  [arXiv:1706.07553 [hep-ph]].

\bibitem{Wu:2016vtq}
  J.~Wu, Y.~R.~Liu, K.~Chen, X.~Liu and S.~L.~Zhu,
  Phys. Rev. D \textbf{97}, no.9, 094015 (2018)
  [arXiv:1605.01134 [hep-ph]].

\bibitem{Wang:2019rdo}
  G.~J.~Wang, L.~Meng and S.~L.~Zhu,
  Phys. Rev. D \textbf{100}, no.9, 096013 (2019)
  [arXiv:1907.05177 [hep-ph]].

\bibitem{Liu:2019zuc}
  M.~S.~Liu, Q.~F.~L\"u, X.~H.~Zhong and Q.~Zhao,
  Phys. Rev. D \textbf{100}, no.1, 016006 (2019)
  [arXiv:1901.02564 [hep-ph]].

\bibitem{Faustov:2020qfm}
  R.~N.~Faustov, V.~O.~Galkin and E.~M.~Savchenko,
  [arXiv:2009.13237 [hep-ph]].

\bibitem{Lu:2020cns}
  Q.~F.~L\"u, D.~Y.~Chen and Y.~B.~Dong,
  Eur. Phys. J. C \textbf{80}, no.9, 871 (2020)
  [arXiv:2006.14445 [hep-ph]].

\bibitem{Heupel:2012ua}
  W.~Heupel, G.~Eichmann and C.~S.~Fischer,
  Phys. Lett. B \textbf{718}, 545-549 (2012)
  [arXiv:1206.5129 [hep-ph]].

\bibitem{Weng:2020jao}
  X.~Z.~Weng, X.~L.~Chen, W.~Z.~Deng and S.~L.~Zhu,
  [arXiv:2010.05163 [hep-ph]].

\bibitem{Berezhnoy:2011xn}
  A.~V.~Berezhnoy, A.~V.~Luchinsky and A.~A.~Novoselov,
  Phys. Rev. D \textbf{86}, 034004 (2012)
  [arXiv:1111.1867 [hep-ph]].

\bibitem{Karliner:2016zzc}
  M.~Karliner, S.~Nussinov and J.~L.~Rosner,
  Phys. Rev. D \textbf{95}, no.3, 034011 (2017)
  [arXiv:1611.00348 [hep-ph]].

\bibitem{Berezhnoy:2011xy}
  A.~V.~Berezhnoy, A.~K.~Likhoded, A.~V.~Luchinsky and A.~A.~Novoselov,
  Phys. Rev. D \textbf{84}, 094023 (2011)
  [arXiv:1101.5881 [hep-ph]].

\bibitem{Feng:2020riv}
  F.~Feng, Y.~Huang, Y.~Jia, W.~L.~Sang, X.~Xiong and J.~Y.~Zhang,
  [arXiv:2009.08450 [hep-ph]].

\bibitem{Ma:2020kwb}
  Y.~Q.~Ma and H.~F.~Zhang,
  [arXiv:2009.08376 [hep-ph]].

\bibitem{Karliner:2020dta}
  M.~Karliner and J.~L.~Rosner,
  [arXiv:2009.04429 [hep-ph]].

\bibitem{Wang:2020wrp}
  J.~Z.~Wang, D.~Y.~Chen, X.~Liu and T.~Matsuki,
  [arXiv:2008.07430 [hep-ph]].


\bibitem{Giron:2020wpx}
  J.~F.~Giron and R.~F.~Lebed,
  Phys. Rev. D \textbf{102}, no.7, 074003 (2020)
  [arXiv:2008.01631 [hep-ph]].

\bibitem{Chao:2020dml}
  K.~T.~Chao and S.~L.~Zhu,
  Sci. Bull. \textbf{65}, 1952-1953 (2020).

\bibitem{Maiani:2020pur}
  L.~Maiani,
  [arXiv:2008.01637 [hep-ph]].

\bibitem{Richard:2020hdw}
  J.~M.~Richard,
  [arXiv:2008.01962 [hep-ph]].

\bibitem{Zhu:2020xni}
  R.~Zhu,
  [arXiv:2010.09082 [hep-ph]].

\bibitem{Guo:2020pvt}
  Z.~H.~Guo and J.~A.~Oller,
  [arXiv:2011.00978 [hep-ph]].

\bibitem{Maciula:2020wri}
  R.~Maciu\l{}a, W.~Sch\"afer and A.~Szczurek,
  [arXiv:2009.02100 [hep-ph]].

\bibitem{Zhu:2020snb}
  J.~W.~Zhu, X.~D.~Guo, R.~Y.~Zhang, W.~G.~Ma and X.~Q.~Li,
  [arXiv:2011.07799 [hep-ph]].

\bibitem{Eichmann:2020oqt}
  G.~Eichmann, C.~S.~Fischer, W.~Heupel, N.~Santowsky and P.~C.~Wallbott,
  Few Body Syst. \textbf{61}, no.4, 38 (2020).

\bibitem{Gong:2020bmg}
  C.~Gong, M.~C.~Du, B.~Zhou, Q.~Zhao and X.~H.~Zhong,
  [arXiv:2011.11374 [hep-ph]].

\bibitem{Becchi:2020uvq}
  C.~Becchi, J.~Ferretti, A.~Giachino, L.~Maiani and E.~Santopinto,
  Phys. Lett. B \textbf{811}, 135952 (2020).

\bibitem{Dong:2020nwy}
  X.~K.~Dong, V.~Baru, F.~K.~Guo, C.~Hanhart and A.~Nefediev,
  [arXiv:2009.07795 [hep-ph]].

\bibitem{Chen:2016jxd}
  W.~Chen, H.~X.~Chen, X.~Liu, T.~G.~Steele and S.~L.~Zhu,
  Phys. Lett. B \textbf{773}, 247-251 (2017)
  [arXiv:1605.01647 [hep-ph]].

\bibitem{Wang:2017jtz}
  Z.~G.~Wang,
  Eur. Phys. J. C \textbf{77}, no.7, 432 (2017)
  [arXiv:1701.04285 [hep-ph]].

\bibitem{Chen:2018cqz}
  W.~Chen, H.~X.~Chen, X.~Liu, T.~G.~Steele and S.~L.~Zhu,
  EPJ Web Conf. \textbf{182}, 02028 (2018)
  [arXiv:1803.02522 [hep-ph]].

\bibitem{Wang:2018poa}
  Z.~G.~Wang and Z.~Y.~Di,
  Acta Phys. Polon. B \textbf{50}, 1335 (2019)
  [arXiv:1807.08520 [hep-ph]].

\bibitem{Zhang:2020xtb}
J.~R.~Zhang,
Phys. Rev. D \textbf{103}, no.1, 014018 (2021)
[arXiv:2010.07719 [hep-ph]].

\bibitem{Wang:2020dlo}
Z.~G.~Wang,
Int. J. Mod. Phys. A \textbf{36}, 2150014 (2021)
[arXiv:2009.05371 [hep-ph]].

\bibitem{Albuquerque:2020hio}
  R.~M.~Albuquerque, S.~Narison, A.~Rabemananjara, D.~Rabetiarivony and G.~Randriamanatrika,
  Phys. Rev. D \textbf{102}, no.9, 094001 (2020)
  [arXiv:2008.01569 [hep-ph]].

\bibitem{Chen:2020xwe}
  H.~X.~Chen, W.~Chen, X.~Liu and S.~L.~Zhu,
  Sci. Bull. \textbf{65}, 1994-2000 (2020)
  [arXiv:2006.16027 [hep-ph]].

\bibitem{Wan:2020fsk}
  B.~D.~Wan and C.~F.~Qiao,
  [arXiv:2012.00454 [hep-ph]].

\bibitem{Latorre:1985uy}
  J.~I.~Latorre and P.~Pascual,
  J. Phys. G \textbf{11}, L231 (1985).

\bibitem{Narison:1986vw}
  S.~Narison,
  Phys. Lett. B \textbf{175}, 88 (1986).

\bibitem{Wang:2006ri}
  Z.~G.~Wang,
  Nucl. Phys. A \textbf{791}, 106-116 (2007)
  [arXiv:hep-ph/0610171 [hep-ph]].

\bibitem{Wang:2015nwa}
  Z.~G.~Wang,
  Int. J. Mod. Phys. A \textbf{30}, no.30, 1550168 (2015)
  [arXiv:1502.01459 [hep-ph]].

\bibitem{Tang:2016pcf}
  L.~Tang and C.~F.~Qiao,
  Eur. Phys. J. C \textbf{76}, no.10, 558 (2016)
  [arXiv:1603.04761 [hep-ph]].

\bibitem{Tang:2019nwv}
  L.~Tang, B.~D.~Wan, K.~Maltman and C.~F.~Qiao,
  Phys. Rev. D \textbf{101}, no.9, 094032 (2020)
 [arXiv:1911.10951 [hep-ph]].

\bibitem{Shifman}
  M.A. Shifman, A.I. Vainshtein and V.I. Zakharov,
  Nucl. Phys. {\bf B147}, 385 (1979); ibid, Nucl. Phys. {\bf B147},
  448 (1979).

\bibitem{Reinders:1984sr}
  L.~J.~Reinders, H.~Rubinstein and S.~Yazaki,
  Phys.\ Rept.\  {\bf 127}, 1 (1985).

\bibitem{Narison:1989aq}
  S.~Narison,
  World Sci.\ Lect.\ Notes Phys.\  {\bf 26} (1989) 1.

\bibitem{P.Col}
  P. Colangelo and A. Khodjamirian, in {\it At the frontier of
  particle physics / Handbook of QCD}, edited by M. Shifman (World
  Scientific, Singapore, 2001), arXiv:hep-ph/0010175.

\bibitem{Wang:2013vex}
  Z.~G.~Wang and T.~Huang,
  Phys.\ Rev.\ D {\bf 89}, 054019 (2014).

\bibitem{Albuquerque:2013ija}
  R.~M.~Albuquerque,
  arXiv:1306.4671 [hep-ph].

\bibitem{Matheus:2006xi}
  R.~D'E.~Matheus, S.~Narison, M.~Nielsen and J.~M.~Richard,
  Phys.\ Rev.\ D {\bf 75}, 014005 (2007).

\bibitem{Finazzo:2011he}
  S.~I.~Finazzo, M.~Nielsen and X.~Liu,
  Phys.\ Lett.\ B {\bf 701}, 101 (2011)
  [arXiv:1102.2347 [hep-ph]].

\bibitem{Qiao:2013raa}
  C.~F.~Qiao and L.~Tang,
  Eur. Phys. J. C \textbf{74}, no.10, 3122 (2014)
  [arXiv:1307.6654 [hep-ph]].

\bibitem{Qiao:2013dda}
  C.~F.~Qiao and L.~Tang,
  Eur. Phys. J. C \textbf{74}, 2810 (2014).

\end{thebibliography}
\end{document}